\documentclass[fleqn,usenatbib]{mnras}

\usepackage{newtxtext,newtxmath}

\usepackage[T1]{fontenc}

\DeclareRobustCommand{\VAN}[3]{#2}
\let\VANthebibliography\thebibliography
\def\thebibliography{\DeclareRobustCommand{\VAN}[3]{##3}\VANthebibliography}

\usepackage{graphicx}	
\usepackage{amsmath}	
\usepackage{subcaption}
\usepackage{xcolor}
\usepackage{longtable}
\usepackage{natbib}
\usepackage{gensymb}

\title{Radio Frequency Interference Detection Using Efficient Multi-Scale Convolutional Attention UNet}
\author[Fei Gu et al.]{
Fei Gu,$^{1}$
Longfei Hao,$^{2}$\thanks{E-mail:haolongfei@ynao.ac.cn}
Bo Liang,$^{1}$\thanks{E-mail: liangbo@kust.edu.cn}
Song Feng,$^{1}$
Shoulin Wei,$^{1}$
Wei Dai,$^{1}$
Yonghua Xu,$^{2}$
\newauthor
Zhixuan Li,$^{2}$
Yihang Dao$^{1}$
\\
$^{1}$Faculty of Information Engineering and Automation, Kunming University of Science and Technology, Kunming 650500,People's Republic of China\\
$^{2}$Yunnan Observatories, Chinese Academy of Science, Kunming, 650000, Yunnan, China
}

\date{Accepted XXX. Received YYY; in original form ZZZ}

\pubyear{2015}

\begin{document}
\label{firstpage}
\pagerange{\pageref{firstpage}--\pageref{lastpage}}
\maketitle

\begin{abstract}

Studying the universe through radio telescope observation is crucial. However, radio telescopes capture not only signals from the universe but also various interfering signals, known as Radio Frequency Interference (RFI). The presence of RFI can significantly impact data analysis. Ensuring the accuracy, reliability, and scientific integrity of research findings by detecting and mitigating or eliminating RFI in observational data, presents a persistent challenge in radio astronomy. In this study, we proposed a novel deep learning model called EMSCA-UNet for RFI detection. The model employs multi-scale convolutional operations to extract RFI features of various scale sizes. Additionally, an attention mechanism is utilized to assign different weights to the extracted RFI feature maps, enabling the model to focus on vital features for RFI detection. We evaluated the performance of the model using real data observed from the 40-meter radio telescope at Yunnan Observatory. Furthermore, we compared our results to other models, including U-Net, RFI-Net, and R-Net, using four commonly employed evaluation metrics: precision, recall, F1 score, and IoU. The results demonstrate that our model outperforms the other models on all evaluation metrics, achieving an average improvement of approximately 5\% compared to U-Net. Our model not only enhances the accuracy and comprehensiveness of RFI detection but also provides more detailed edge detection while minimizing the loss of useful signals.
\end{abstract}




\section{Introduction}

Being essential for studying the universe, radio telescopes capture radio signals from the cosmos, invariably confronting Radio Frequency Interference (RFI) in the process.In light of the rapid advancement of human communication technology,signals generated by human activities are increasingly occupying a larger portion of the frequency bands. Consequently, the interference to radio observation equipment within these frequency bands has intensified and substantially compromised the quality of radio astronomy observation data \citep{yan2021radio}. Therefore, in order to ensure the accuracy and validity of radio astronomy research, it is crucial to accurately detect RFI from complex radio observations.

Traditional RFI detection methods are mainly Singular Vector Decomposition \citep{offringa2010post}, Principle Component Analysis \citep{ Zhao2013WindSatRI}, CUMSUM \citep{baan2004radio} and SUMTHRESHOLD \citep{offringa2010post}, which have been widely used in various radio data processing pipelines \citep{akeret2017hide,offringa2010lofar,offringa2010post,peck2013serpent}. However, these traditional methods are largely influenced by human empirical factors, increasing the need for manual intervention and greatly reducing the efficiency of data processing. In addition, some new Bayesian-based statistical methods have been used for RFI mitigation in more recent studies\citep{leeney2023bayesian,finlay2023trajectory}.

In order to overcome the limitations of the above methods, the use of deep learning techniques based on computer vision to detect RFI has also been gradually explored. The U-Net \citep{ronneberger2015u} model was initially used by \cite{akeret2017radio} for RFI detection. After testing on simulated single-dish data, the U-Net architecture demonstrated superior results to traditional methods.\cite{kerrigan2019optimizing}achieved promising outcomes by combining amplitude and phase information and utilizing a fully convolutional neural network to identify RFI in HERA\citep{deboer2017hydrogen} data. Inspired by residual networks\citep{he2016deep}, \cite{yang2020deep}proposed the RFI-Net model by extending the U-Net model. This model outperformed U-Net on several datasets, including FAST. In more recent studies, transfer learning\citep{tan2018survey} has also been used for RFI detection. The R-Net model, proposed by \cite{vafaei2020deep}, was first trained on simulated data and then fine-tuned on real data, achieving superior performance compared to U-Net.

Most existing deep learning methods have demonstrated promising results in simulating data. However, when applied to detect RFI in the real observation data from the Yunnan Observatory's 40-meter radio telescope, these models often misclassify non-RFI as RFI and miss certain instances of RFI that go undetected, necessitating additional manual inspection. Because compared to the simple RFI types in simulated data, RFI in real data is usually more complex. To address these challenges and enhance RFI detection in real data, we propose a novel deep learning model that combines the advantages of attention mechanism and multi-scale convolutional neural networks. 

In contrast to the aforementioned deep learning models, we have replaced the single-sized small convolution kernel with multiple-sized large convolution kernels, and integrated an attention mechanism into the convolutional neural network. While convolutional neural networks are effective at extracting local information, they often overlook global information. However, the inclusion of attention mechanisms assists the model in considering global information. Therefore, the combination of these elements allows our model to simultaneously handle both local and global patterns, resulting in improved detection of RFI in real data. Our model utilizes multi-scale convolutional operations to extract informative RFI features and generate corresponding feature maps. Subsequently, through the attention mechanism, the model focuses on the relevant features crucial for RFI detection, rather than processing all the information in the feature maps, thereby improving the efficiency of detection.

The structure of this paper is as follows: In Section~\ref{sec:data},we describe the telescope used for data acquisition and provide details of the experimental data. In Section~\ref{sec:method}, the composition and functions of each module in our model are presented. Next, the details of our experiments are described in Section~\ref{sec:experiments}. The evaluation of the experimental results and the analysis and discussion of their reasons are addressed in Section~\ref{sec:results and discussion}. Finally, our conclusions are provided in Section~\ref{sec:conclusion}.

\section{Data }
\label{sec:data}

\begin{figure}
 \includegraphics[width=\columnwidth]{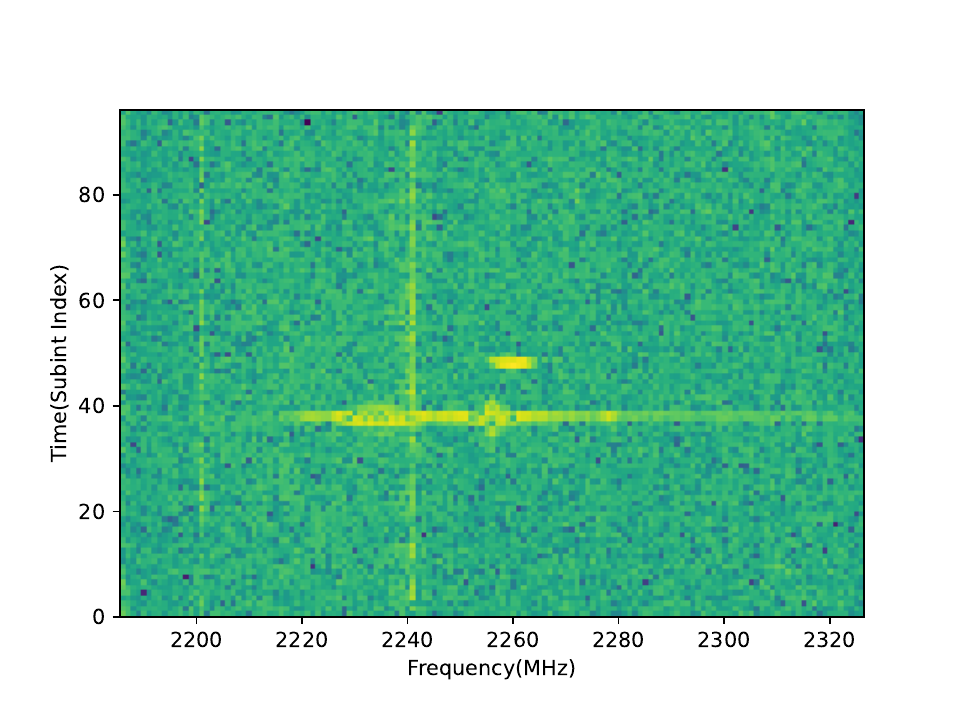}
 \caption{An example of the actual observational data of PSR J0032+5434 collected by the 40-meter radio telescope at Yunnan Observatory. The time axis has 91 subintegrations of 10 s each, while the frequency axis contains 141 frequency channels (2185.5-2326.5MHz). This panel clearly shows several types of typical RFI, including two long-duration narrowband RFI signals spanning multiple subintegrations at 2201MHz and 2241MHz. Additionally, wideband RFI spans multiple frequency channels observed during the 36th to 39th subintegrations. Another relatively short-duration RFI is observed around 2260MHz.}
 \label{fig:time-frequency image}
\end{figure}

The experimental data were obtained through pulsar timing observations conducted using the Yunnan Observatory's 40-meter radio telescope, situated in the southwestern region of China (25\degree 01'38'' N, 102\degree 47'45'' E). This telescope is equipped with a room-temperature S/X receiver and a cooled circularly polarized C-band receiver, which was installed in 2016. The system temperatures for the Yunnan Observatory's 40-meter radio telescope in the S and C bands are 70 K and 30 K, respectively\citep{xu2023interstellar}. Specifically designed for pulsar observation, the telescope operates within the S band (2150 MHz-2450 MHz), with a center frequency of 2256 MHz. Given the typically weak nature of pulsar signals, these are folded over a specified integration time (e.g., 30 seconds, also termed "subintegration" or "subint") aligning with the pulsar signal period for attaining clearer pulse profiles. The folded pulsar data is stored in the standard PSRFITS format\citep{hotan2004psrchive} for further analysis. In order to capture the characteristics of RFI comprehensively, we selected observation data from September 2016 to October 2022, which includes data from three different pulsars: PSR J0332+5434, J00358+5413, and J0437-4715. We select data with the number of subintegration greater than 90 and the number of frequency channels greater than 128 for analysis. Fig.~\ref{fig:time-frequency image} displays the time-frequency image of the actual observation data from PSR J0032+5434, where the typical RFI can be observed to be much brighter than the background noise or astronomical signals.

All the corresponding labels for the data were generated using the aoflagger algorithm\citep{offringa2023interference}. Diverse strategies of aoflagger exert varied effects on identifying  different types of RFI; selecting appropriate flagging strategies for respective RFI can typically yield favorable results. Consequently,when creating the ground truth, we initially employ various flagging strategies of Aoflagger to flag the data, followed by manual verification of the labels' accuracy. Our study's dataset encompasses 1384 data samples. Due to the relatively small size of the dataset, in order to reduce overfitting and enhance the generalizability of the model, we performed random cropping and random horizontal flipping as data augmentation operations on the training data. The size of the time-frequency images used for model training and evaluation is 256×256.

\section{Method}
\label{sec:method}

\begin{table*}
\centering
\caption{Stem block detailed structure. In all subsequent detailed structure descriptions, we use Pytorch's data dimensions (B, C, H, W) to describe the size of the input and output data, in the table, B is the Batch size, which represents the number of inputs sent to the model at a time, Conv is the common convolution operation, and BN stands for the batch normalization, which can make the training process more stable. GELU is an activation function based on a Gaussian function, which helps to increase the expressive power of the model and provides smoother gradients through its nonlinear properties. Short cut is a short connection also known as a residual connection. Add is an element-level addition operation that sums the results from short cut and conv2.}
\label{tab:Stem block}
\begin{tabular}{|c|c|c|c|c|}
\hline
Layer     & Operation                     & Input size                                  & Output size      & Kernel size \\ \hline
short cut & Conv                          & B×3×256×256                                 & B×64×256×256     & 1×1         \\ \hline
conv1     & Conv+BN+GELU                  & B×3×256×256                                 & B×32×256×256     & 3×3         \\ \hline
conv2     & Conv+BN                       & B×32×256×256                                & B×64×256×256     & 3×3         \\ \hline
add       & Add                           & \begin{tabular}[c]{@{}l@{}}B×64×256×256(from short cut results),\\ B×64×256×256(from conv2 results)\end{tabular} & B×64×256×256 & -           \\ \hline
\end{tabular}
\end{table*}

{
\renewcommand{\arraystretch}{0.8} 
\begin{table*}
\centering
\caption{Detailed structure of EMSCA-UNet. TransposeConv is the transpose convolution, sometimes called the inverse convolution, which is used to upsample to gradually recover the resolution of the image. In the final output layer, we use a sigmoid activation function to generate the probability values predicted by the model. The detailed structure of the StemConv Block, the ECA Block, and the MSCA Block are shown in Table~\ref{tab:Stem block}, ~\ref{tab:ECA block}, and ~\ref{tab:MSCA block}.}
\label{tab:EMSCA-UNet}
\begin{tabular}{|c|c|c|c|c|}
\hline
Layer         & Operation             & Input size                                                                                                           & Output size   & Kernel size \\ \hline
stemconv      & StemConv Block        & B×3×256×256                                                                                                          & B×64×256×256  & -           \\ \hline
eca1          & ECA Block             & B×64×256×256                                                                                                         & B×64×256×256  & -           \\ \hline
msca1         & MSCA Block            & B×64×256×256                                                                                                         & B×64×256×256  & -           \\ \hline
down sample 1 & Conv+BN+GELU          & B×64×256×256                                                                                                         & B×128×128×128 & 2×2         \\ \hline
eca2          & ECA Block             & B×128×128×128                                                                                                        & B×128×128×128 & -           \\ \hline
msca2         & MSCA Block            & B×128×128×128                                                                                                        & B×128×128×128 & -           \\ \hline
down sample 2 & Conv+BN+GELU          & B×128×128×128                                                                                                        & B×256×64×64   & 2×2         \\ \hline
eca3          & ECA Block             & B×256×64×64                                                                                                          & B×256×64×64   & -           \\ \hline
msca3         & MSCA Block            & B×256×64×64                                                                                                          & B×256×64×64   & -           \\ \hline
down sample 3 & Conv+BN+GELU          & B×256×64×64                                                                                                          & B×512×32×32   & 2×2         \\ \hline
eca4          & ECA Block             & B×512×32×32                                                                                                          & B×512×32×32   & -           \\ \hline
msca4         & MSCA Block            & B×512×32×32                                                                                                          & B×512×32×32   & -           \\ \hline
down sample 4 & Conv+BN+GELU          & B×512×32×32                                                                                                          & B×1024×16×16  & 2×2         \\ \hline
eca5          & ECA Block             & B×1024×16×16                                                                                                         & B×1024×16×16  & -           \\ \hline
msca5         & MSCA Block            & B×1024×16×16                                                                                                         & B×1024×16×16  & -           \\ \hline
up sample 1   & TransposeConv+BN+GELU & B×1024×16×16                                                                                                         & B×512×32×32   & 2×2         \\ \hline
concat  1     & Concatenate           & \begin{tabular}[c]{@{}l@{}}B×512×32×32(from msca4 results),\\ B×512×32×32(from up sample 1 results)\end{tabular}     & B×1024×32×32  & -           \\ \hline
eca6          & ECA Block             & B×1024×32×32                                                                                                         & B×1024×32×32  & -           \\ \hline
conv1         & Conv                  & B×1024×32×32                                                                                                         & B×512×32×32   & 1×1         \\ \hline
msca6         & MSCA Block            & B×512×32×32                                                                                                          & B×512×32×32   & -           \\ \hline
up sample 2   & TransposeConv+BN+GELU & B×512×32×32                                                                                                          & B×256×64×64   & 2×2         \\ \hline
concat  2     & Concatenate           & \begin{tabular}[c]{@{}l@{}}B×256×64×64(from msca3 results),\\ B×256×64×64(from up sample 2 results)\end{tabular}     & B×512×64×64   & -           \\ \hline
eca7          & ECA Block             & B×512×64×64                                                                                                          & B×512×64×64   & -           \\ \hline
conv2         & Conv                  & B×512×64×64                                                                                                          & B×256×64×64   & 1×1         \\ \hline
msca7         & MSCA Block            & B×256×64×64                                                                                                          & B×256×64×64   & -           \\ \hline
up sample 3   & TransposeConv+BN+GELU & B×256×64×64                                                                                                          & B×128×128×128 & 2×2         \\ \hline
concat  3     & Concatenate           & \begin{tabular}[c]{@{}l@{}}B×128×128×128(from msca2 results),\\ B×128×128×128(from up sample 3 results)\end{tabular} & B×256×128×128 & -           \\ \hline
eca8          & ECA Block             & B×256×128×128                                                                                                        & B×256×128×128 & -           \\ \hline
conv3         & Conv                  & B×256×128×128                                                                                                        & B×128×128×128 & 1×1         \\ \hline
msca8         & MSCA Block            & B×128×128×128                                                                                                        & B×128×128×128 & -           \\ \hline
up sample 4   & TransposeConv+BN+GELU & B×128×128×128                                                                                                        & B×64×256×256  & 2×2         \\ \hline
concat  4     & Concatenate           & \begin{tabular}[c]{@{}l@{}}B×64×256×256(from msca1 results),\\ B×64×256×256(from up sample 4 results)\end{tabular}   & B×128×256×256 & -           \\ \hline
eca9          & ECA Block             & B×128×256×256                                                                                                        & B×128×256×256 & -           \\ \hline
conv4         & Conv                  & B×128×256×256                                                                                                        & B×64×256×256  & 1×1         \\ \hline
msca9         & MSCA Block            & B×64×256×256                                                                                                         & B×64×256×256  & -           \\ \hline
output        & Conv+BN+Sigmoid       & B×64×256×256                                                                                                         & B×1×256×256   & -           \\ \hline
\end{tabular}
\end{table*}
}

\begin{figure*}
	\includegraphics[width=\textwidth]{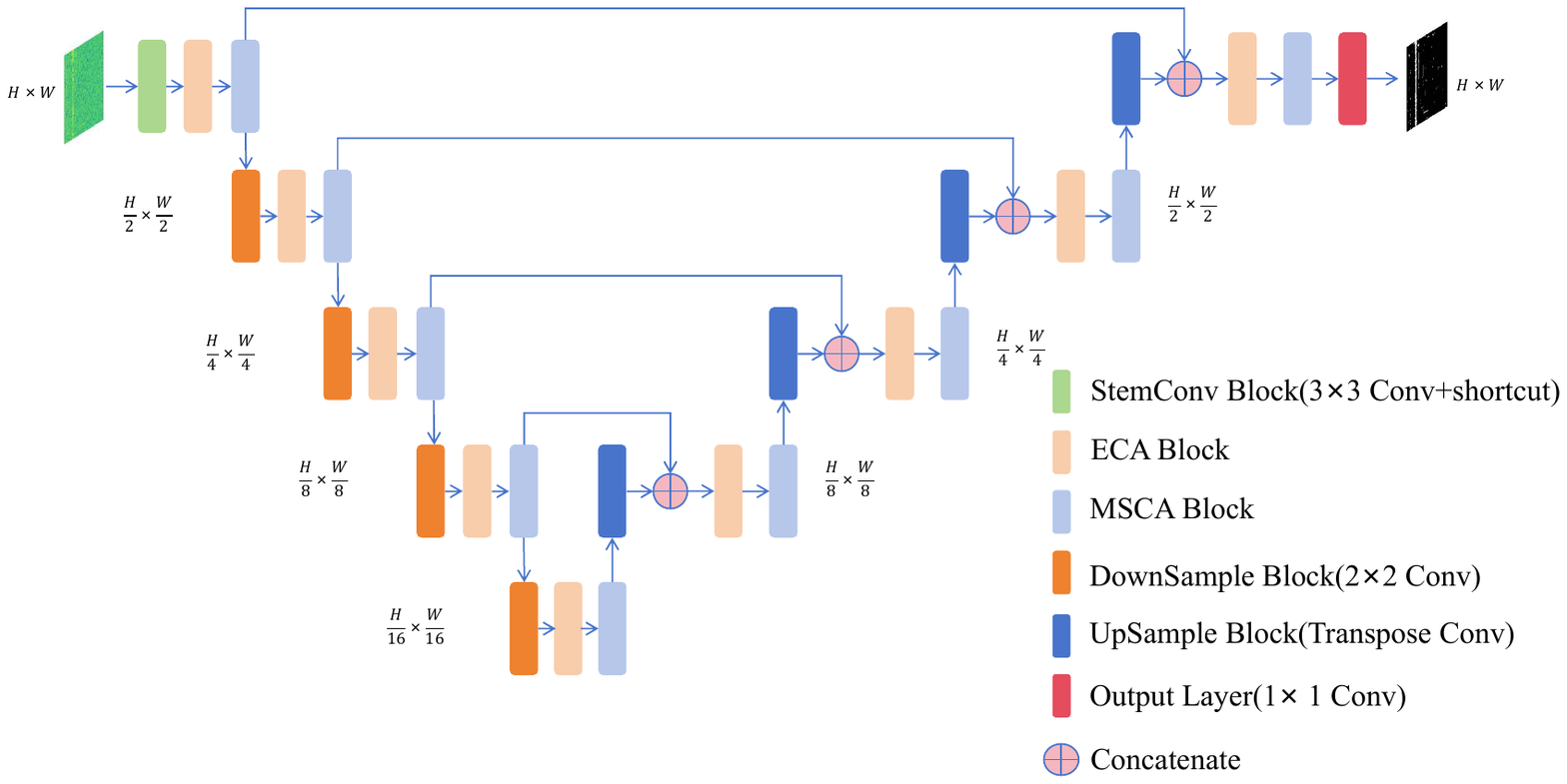}
    \caption{Schematic diagram of the EMSCA-UNet structure.}
    \label{fig:EMSCA-UNet}
\end{figure*}

To address the limitations of existing methods, we propose a new deep learning model for RFI detection, which we refer to as EMSCA-UNet. The network schematic diagram is shown in Fig.~\ref{fig:EMSCA-UNet}, and the detailed structure can be found in Table~\ref{tab:EMSCA-UNet}. Next, we will provide a detailed overview of the overall architecture of our proposed model and discuss the specifics of each module that constitutes this model.

\subsection{Architecture of EMSCA-UNet}
\label{sec:Architecture of EMSCA-UNet}

In constructing the backbone network of our model, we draw inspiration from the multi-scale convolutional attention mechanism (MSCA) proposed by \cite{guo2022segnext} and the efficient channel attention mechanism (ECA) introduced by \cite{wang2020eca}. The combination of these two mechanisms provides the ability for channel weight adaptation and multi-scale feature processing in RFI detection. Our model is based on a clear and effective encoder-decoder architecture. For the input data, we initially apply a StemConv block (detailed in Table~\ref{tab:Stem block}), consisting of a dual-layer 3×3 convolution and residual connection \citep{he2016deep}, for preliminary processing. This design not only prevents gradient vanishing and stabilizes the training process but also extracts low-level features of RFI from the images, providing richer feature information for subsequent layers. In the subsequent encoding and decoding stages, ECA blocks are used to learn the correlations between different channels, adaptively adjusting the weight of each channel to improve the performance and efficiency of the network. Then, the channels with different weights are input into MSCA blocks to extract multi-scale features of RFI. Throughout the encoding and decoding process, we stack a series of these ECA and MSCA blocks to construct a U-shaped network structure \citep{ronneberger2015u}.

In the encoder part, we replaced the previous max pooling down-sampling method with 2×2 convolutional layers, which reduces both spatial and semantic information loss while decreasing the computational burden. In the decoder part, we restored the original image resolution gradually by using transpose convolution to preserve as much image detail as possible. High-resolution feature maps contain rich details and low-level information, but they may have negative impacts on RFI detection. On the other hand, low-resolution feature maps have higher semantic information but lack specific spatial details. Therefore, in the decoding process, we adopted skip connections similar to Unet to integrate the low-level and high-level RFI features \citep{ronneberger2015u}. Through skip connections, we can combine contextual information and multi-scale information, maintain the spatial structure and detail information of the image, compensate for the spatial information loss caused by downsampling, and improve the expressiveness and detection accuracy of the network. Since skip connections double the number of channels, we used 1×1 convolutional layers to re-model and fuse the cross-channel features in order to reduce the number of channels. In the final output layer, we used a 1×1 convolutional layer to reduce the number of channels to 1. Then, we applied the sigmoid activation function to map all results to a probability range of 0 to 1. By setting the threshold to 0.5, we obtained the output segmentation mask. To prevent overfitting, we also introduced the droppath mechanism \citep{huang2016deep} to enhance the model's generalization ability. In this study, we set the droppath rate to 0.2. As for our model architecture, the use of a U-shaped structure is something that has been done previously; however, those approaches solely employed single-scale small convolutional kernels in their convolutional neural networks. We have replaced these small, single-scale convolutional kernels with larger, multi-scale ones and have integrated attention mechanisms into the network, which is a novelty not found in the methods we compared with. All hyperparameters in this network architecture were carefully tuned and optimized through extensive experimentation.

\subsection{ECA block}


\begin{figure}
	\makebox[\columnwidth][l]{\hspace{-1cm}\includegraphics[width=1.25\columnwidth]{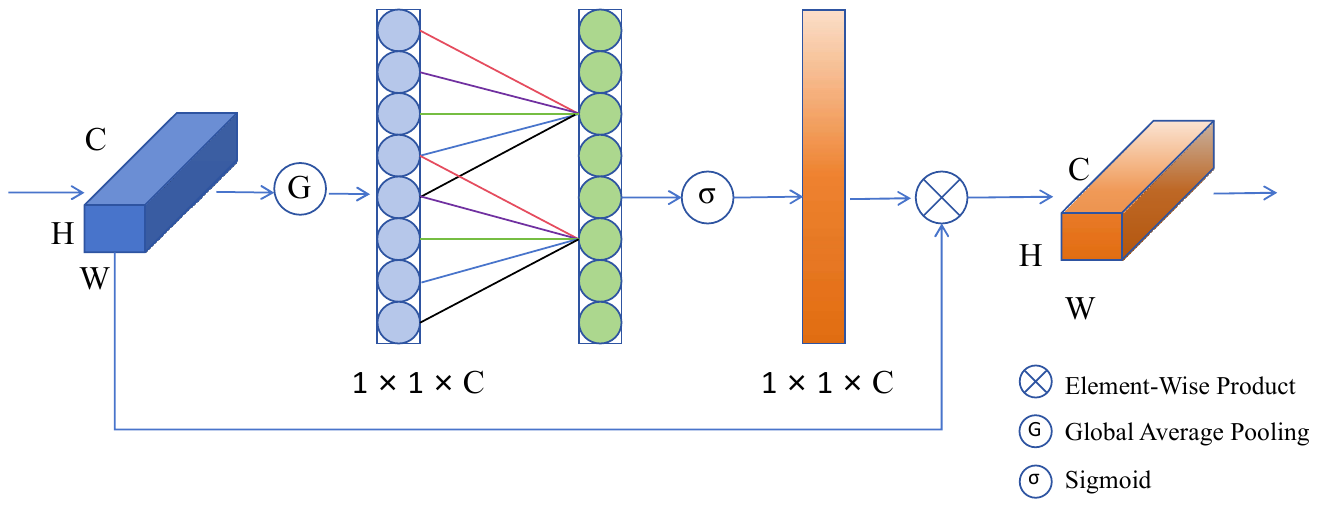}}
    \caption{Schematic of the Efficient Channel Attention (ECA) module. The aggregated features are first obtained using global average pooling (GAP), and then ECA generates the channel weights by performing a fast one-dimensional convolution with convolution kernel size k, where k is adaptively determined by mapping the channel dimension C.}
    \label{fig:ECA Block}
\end{figure}

\begin{table*}
\centering
\caption{Detailed structure of the ECA block. C represents the number of channels, H represents the height of the image data, W represents the width of the image data, and Element-Wise Product represents the element-wise multiplication operation. Reshape is used to change the dimensionality of the data to ensure dimension compatibility for subsequent operations.}
\label{tab:ECA block}
\begin{tabular}{|c|c|c|c|c|}
\hline
Layer    & Operation                        & Input size                                             & Output size & Kernel size                 \\ \hline
pool     & Adaptive Average Pooling+Reshape & B×C×H×W                                                & B×1×C       & -                           \\ \hline
conv     & 1D Conv                          & B×1×C                                                  & B×1×C       & kernel\_size=~(\ref{eq:kernalsize}) \\ \hline
sigmoid  & Sigmoid activation+Reshape       & B×1×C                                                  & B×C×1×1     & -                           \\ \hline
multiply & Element-Wise Product             & \begin{tabular}[c]{@{}l@{}}B×C×H×W(from input results),\\ B×C×1×1(from sigmoid results)\end{tabular} & B×C×H×W     & -                           \\ \hline
\end{tabular}
\end{table*}

When observing an image, we typically pay more attention to the parts that interest us. Similarly, attention mechanisms assist neural networks in better utilizing input information. In our network, after the RFI image is processed by convolutional layers, multiple feature maps are generated at different stages. However, not all of these feature maps contribute equally to RFI detection. We should prioritize the feature maps that are more helpful for detecting RFI. Therefore, we employ an efficient Channel Attention mechanism (ECA) to assign different weights to different feature maps, thereby enhancing RFI detection. The schematic diagram of the ECA module is shown in Fig.~\ref{fig:ECA Block}, and its detailed structure can be found in Table~\ref{tab:ECA block}. During the operation of this module on the input data, global average pooling is applied first to aggregate the RFI features extracted by the convolution. Then, a one-dimensional convolution with a kernel size of k is used to generate weights for each feature map. These weights are then mapped to the range of 0 to 1 using the Sigmoid function, representing the learned channel attention \citep{wang2020eca}. Throughout this process, the number of feature maps remains unchanged. The one-dimensional kernel size, k, can be adaptively determined and calculated using Equation ~(\ref{eq:kernalsize}),

\begin{equation}
    k=\left | \frac{\log_{2}{(C)}+b }{\gamma }  \right | _{odd},
	\label{eq:kernalsize}
\end{equation}
$ \left | t \right | _{\text{odd}} $ represents the closest odd number to $t$ and $C$ represents the number of feature maps, i.e., the number of channels. In this paper, we set the values of $\gamma$ and $b$ to 2 and 1, respectively.

\subsection{MSCA block}

\begin{figure*}
	\includegraphics[width=\textwidth]{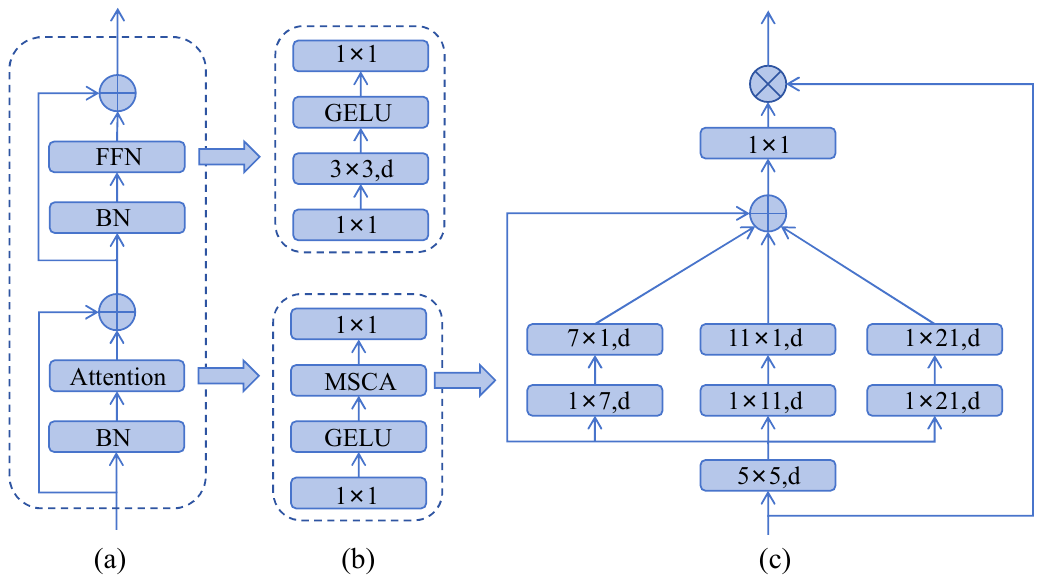}
    \caption{Schematic diagram of MSCA Block structure. In which d represents depth-wise convolution, and k1×k2 represents the size of the convolution kernel. We utilize convolution to extract multi-scale features, and then use them as attention weights to reweight the input to MSCA.}
    \label{fig:EMSCA}
\end{figure*}

The detection requirement for the RFI model is to be able to identify the presence of RFI at the pixel level, which is essentially a binary classification semantic segmentation problem. According to recent research, an efficient semantic segmentation model needs to have the capability of multi-scale information interaction\citep{guo2022segnext}. Given the diverse shapes and features of RFI, the complex RFI patterns necessitate the model to possess the ability to handle multi-scale information.

In response to this requirement, the MSCA block is designed to capture the multi-scale features of RFI. Each MSCA block, as shown in Fig.~\ref{fig:EMSCA}(a), consists of Batch Normalization (BN), an attention module for enhancing key information, and a Feed-Forward Neural Network (FFN). The attention module, as depicted in Fig.~\ref{fig:EMSCA}(b), is composed of a 1×1 convolution, Gaussian Error Linear Unit (GELU) \citep{hendrycks2016gaussian}activation function, and an MSCA module. The FFN is composed of a 1×1 convolution, 3×3 depth convolution, and GELU activation function. As illustrated in Fig.~\ref{fig:EMSCA}(c), the local details and multi-scale information of RFI are fused using the multi-branch depthwise separable convolution, and the information from multiple feature channels is integrated using a 1×1 convolution. The MSCA can be mathematically represented as follows:

\begin{equation}
   A_{m} =Conv_{1\times1 } \left ( \sum_{i=0}^{3}Scale_{i}\left ( \text{DW-Conv}\left ( F \right )  \right )    \right ) ,
	\label{eq:att}
\end{equation}

\begin{equation}
   Out=A_{m} \otimes F.
	\label{eq:out}
\end{equation}

In these two formulas ~(\ref{eq:att}) and ~(\ref{eq:out}), $A_{m} $ represents attention map and $Out$ represents output. $F$ represents input features, $\text{DW-Conv}$ represents depth-separable convolution, $Scale_{i}$,$i\in \left \{ 0,1,2,3 \right \}$  represents the $i$th branch in Fig.~\ref{fig:EMSCA}(c), $Scale_{0}$ is the identity connection, and $\otimes $ represents the element-wise Matrix multiplication operation, the detailed structure of MSCA block is shown in the table~\ref{tab:MSCA block}. We use two depthwise separable strip convolutions in each branch to approximately simulate standard depthwise separable convolutions with large convolution kernels. This is mainly based on two considerations. First, strip convolution is more lightweight and can reduce computational complexity. Secondly, this strip structure has been found to be particularly effective in capturing strip-like features that correspond to certain types of RFI\citep{hou2020strip}. Therefore, using strip convolution can make the model more accurate in identifying such strip-like RFI.

\begin{table*}
\centering
\caption{
Detailed structure of MSCA block. DWconv is a Depthwise separable convolution. Compared with ordinary convolution operations, the number of parameters and operation cost are lower. Clone represents a copy operation, that is, the data is not changed, and it is for subsequent implementation of residual connection. LayerScale is layer normalization, and its function is similar to BN. Because the output of the attention layer and the subsequent Mlp layer in the MSCA block may be very different, so in order to prevent gradient disappearance and gradient explosion, the output of these two places needs to be standardized. Mlp\_Ratio is a hyperparameter used to control the number of input and output channels in the Mlp layer. In our network architecture, except for setting Mlp\_Ratio to 2 in the lowest-level MSCA module, everything else is set is 4.
} 
\label{tab:MSCA block}
\begin{tabular}{|c|c|c|c|c|}
\hline
Layer       & Operation               & Input size                                                                                                                                        & Output size        & Kernel size \\ \hline
conv1       & BN+Conv+GELU            & B×C×H×W                                                                                                                                           & B×C×H×W            & 1×1         \\ \hline
short cut 1 & Clone                   & B×C×H×W                                                                                                                                           & B×C×H×W            & -           \\ \hline
conv2       & DWConv                  & B×C×H×W                                                                                                                                           & B×C×H×W            & 5×5         \\ \hline
conv3       & DWConv                  & B×C×H×W                                                                                                                                           & B×C×H×W            & 1×7         \\ \hline
conv4       & DWConv                  & B×C×H×W                                                                                                                                           & B×C×H×W            & 7×1         \\ \hline
conv5       & DWConv                  & B×C×H×W                                                                                                                                           & B×C×H×W            & 1×11        \\ \hline
conv6       & DWConv                  & B×C×H×W                                                                                                                                           & B×C×H×W            & 11×1        \\ \hline
conv7       & DWConv                  & B×C×H×W                                                                                                                                           & B×C×H×W            & 1×21        \\ \hline
conv8       & DWConv                  & B×C×H×W                                                                                                                                           & B×C×H×W            & 21×1        \\ \hline
conv9       & Conv                    & B×C×H×W                                                                                                                                           & B×C×H×W            & 1×1         \\ \hline
short cut 2 & Clone                   & B×C×H×W                                                                                                                                           & B×C×H×W            & -           \\ \hline
add 1       & Add                     & \begin{tabular}[c]{@{}l@{}}B×C×H×W(from conv2 results),\\ B×C×H×W(from conv4 results),\\ B×C×H×W(from conv6 results),\\ B×C×H×W(from conv8 results)\end{tabular} & B×C×H×W            & -           \\ \hline
multiply    & Element-Wise Product    & \begin{tabular}[c]{@{}l@{}}B×C×H×W(from conv9 results),\\ B×C×H×W(from short cut 2 results)\end{tabular}                                         & B×C×H×W            & -           \\ \hline
conv10      & Conv                    & B×C×H×W                                                                                                                                           & B×C×H×W            & 1×1         \\ \hline
add 2       & Add+LayerScale+DropPath & \begin{tabular}[c]{@{}l@{}}B×C×H×W(from conv10 results),\\ B×C×H×W(from short cut 1 results)\end{tabular}                                        & B×C×H×W            & -           \\ \hline
short cut 3 & Clone                   & B×C×H×W                                                                                                                                           & B×C×H×W            & -           \\ \hline
conv11      & Conv                    & B×C×H×W                                                                                                                                           & B×Mlp\_Ratio×C×H×W & 1×1         \\ \hline
conv12      & DWConv+GELU             & B×Mlp\_Ratio×C×H×W                                                                                                                                & B×Mlp\_Ratio×C×H×W & 3×3         \\ \hline
conv13      & Conv                    & B×Mlp\_Ratio×C×H×W                                                                                                                                & B×C×H×W            & 1×1         \\ \hline
add 3       & Add+LayerScale+DropPath & \begin{tabular}[c]{@{}l@{}}B×C×H×W(from conv13 results),\\ B×C×H×W(from short cut 3 results)\end{tabular}                                        & B×C×H×W            & -           \\ \hline
\end{tabular}
\end{table*}

\section{Experiments}
\label{sec:experiments}

Our experiment was conducted on the Ubuntu 22.04 operating system, using an NVIDIA 3090 graphics card with 24GB of memory. We used Scikit-learn, a machine learning library for Python, to randomly partition all the data into training, validation, and test sets in the ratio of 70\%, 15\%, and 15\%. To ensure the reproducibility of the results, we used the random number seed 3407 on which all subsequent tests were done. The training set was used to train the model, while hyperparameter optimization and model selection were performed with the validation set. Due to computational resource constraints, all hyperparameters in our experiments are manually tuned by the performance on the validation set, since searching for hyperparameters is very time-consuming and requires a lot of computational resources. These hyperparameters written in this paper are optimized. Finally, the test set was used to report the results. We compared several supervised deep learning models that have significant impact in RFI detection, including U-Net \citep{akeret2017radio}, RFI-Net \citep{yang2020deep}, and R-Net \citep{vafaei2020deep}.

In our experiment, we employed the same training and evaluation strategies for all models. The training batch size was set to 8, with 8 samples selected for each iteration. The validation and test batch sizes were set to 1, with one sample selected for testing at a time. The number of epochs was set to 500. For the loss function, we utilized the binary cross-entropy (BCE) loss function built-in in PyTorch. We also tried the BCEWithLogitsLoss with weighted samples but found no significant improvement in model performance. We choose adamW\citep{loshchilov2017decoupled} as the optimizer to train the model, which is a variant of adam but has better generalization performance than adam. Additionally, to optimize the model training process, we applied a dynamic learning rate strategy. We initially set the learning rate to a larger value of 0.001 to enable the model to quickly converge to a better solution. As training progressed, we gradually decreased the learning rate to fine-tune the parameters more accurately and precisely near the optimal solution. This strategy helps prevent large parameter oscillations in the later stages of training, promoting a more stable training process.

In terms of selection of evaluation indicators, in addition to the three key evaluation indicators commonly used in supervised deep learning models for RFI detection in the past: Precision, Recall, and F1 Score, we also added a new evaluation indicator commonly used in semantic segmentation: Intersection over Union (IoU).

The combination of these indicators can comprehensively evaluate the performance of the model from all aspects. The formulas for these indicators are shown in Appendix ~\ref{sec:appendix}.

\label{sec:Experimental details}

\section{RESULTS AND DISCUSSION}
\label{sec:results and discussion}

We will compare the experimental results of our method with other methods through both visual and metric aspects. In Section~\ref{sec:Performance evaluation}, we will evaluate the performance of the model; subsequently, in Section~\ref{sec:Comparative analysis and discussion}, we analyze and discuss the experimental results.

\begin{figure*}
  \centering
  \begin{subfigure}[t]{0.18\textwidth}
    \includegraphics[width=\textwidth]{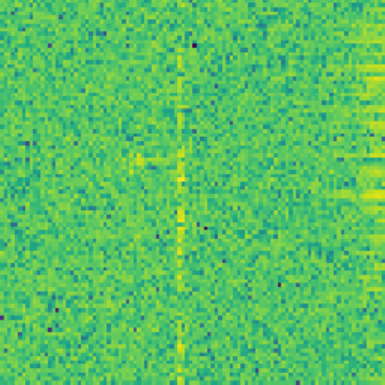}
    \caption{Visibility data}
    \label{fig:1}
  \end{subfigure}\hspace{1em}
  \begin{subfigure}[t]{0.18\textwidth}
    \includegraphics[width=\textwidth]{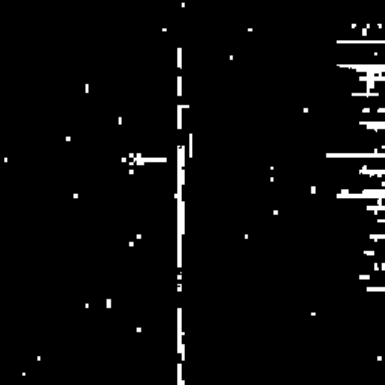}
    \caption{Ours mask}
    \label{fig:2}
  \end{subfigure}\hspace{1em}
  \begin{subfigure}[t]{0.18\textwidth}
    \includegraphics[width=\textwidth]{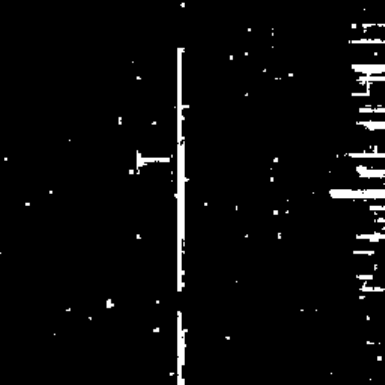}
    \caption{R-Net mask}
    \label{fig:3}
  \end{subfigure}\hspace{1em}
  \begin{subfigure}[t]{0.18\textwidth}
    \includegraphics[width=\textwidth]{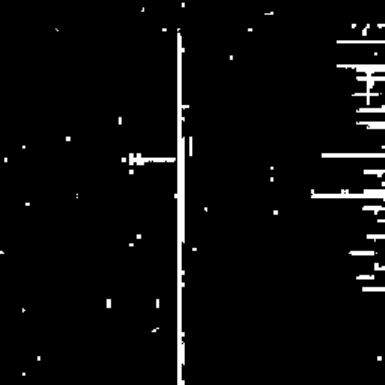}
    \caption{RFI-Net mask}
    \label{fig:4}
  \end{subfigure}\hspace{1em}
  \begin{subfigure}[t]{0.18\textwidth}
    \includegraphics[width=\textwidth]{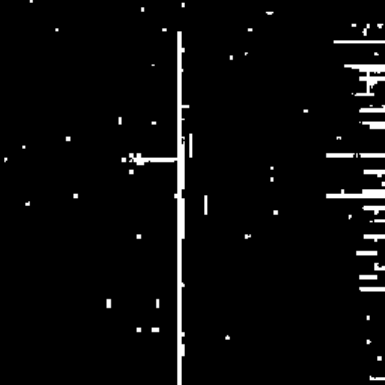}
    \caption{U-Net mask}
    \label{fig:5}
  \end{subfigure}\\
  \begin{subfigure}[t]{0.18\textwidth}
    \includegraphics[width=\textwidth]{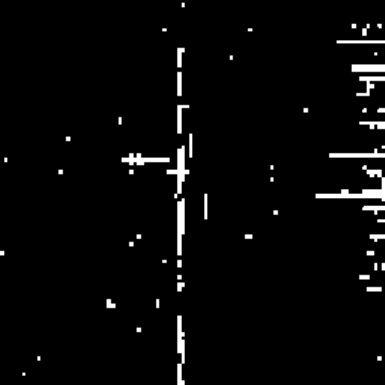}
    \caption{Ground truth}
    \label{fig:6}
  \end{subfigure}\hspace{1em}
  \begin{subfigure}[t]{0.18\textwidth}
    \includegraphics[width=\textwidth]{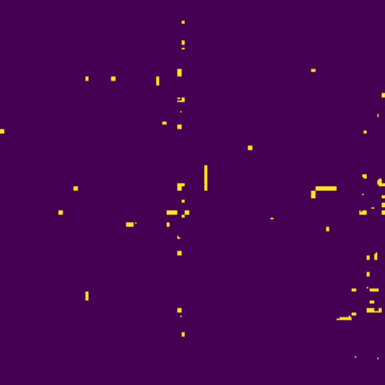}
    \caption{Ours difference map}
    \label{fig:7}
  \end{subfigure}\hspace{1em}
  \begin{subfigure}[t]{0.18\textwidth}
    \includegraphics[width=\textwidth]{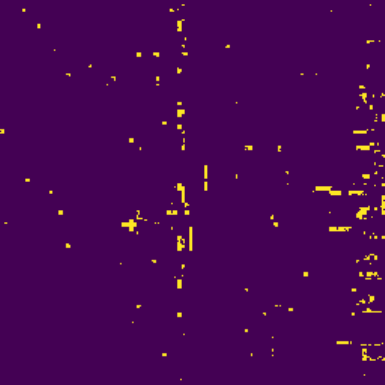}
    \caption{R-Net difference map}
    \label{fig:8}
  \end{subfigure}\hspace{1em}
  \begin{subfigure}[t]{0.18\textwidth}
    \includegraphics[width=\textwidth]{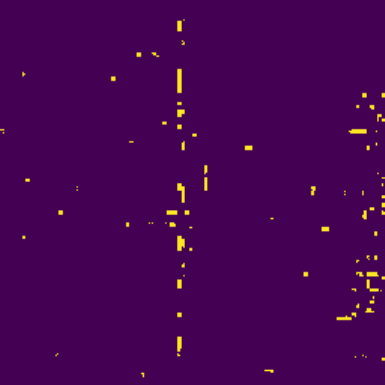}
    \caption{RFI-Net difference map}
    \label{fig:9}
  \end{subfigure}\hspace{1em}
  \begin{subfigure}[t]{0.18\textwidth}
    \includegraphics[width=\textwidth]{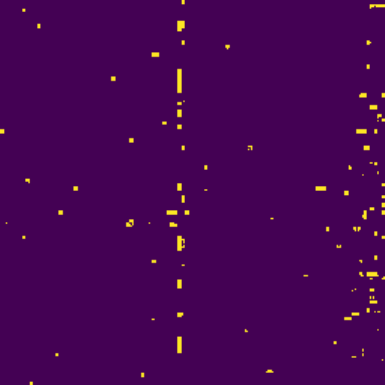}
    \caption{U-Net difference map}
    \label{fig:10}
  \end{subfigure}
   \begin{subfigure}[t]{0.18\textwidth}
    \includegraphics[width=\textwidth]{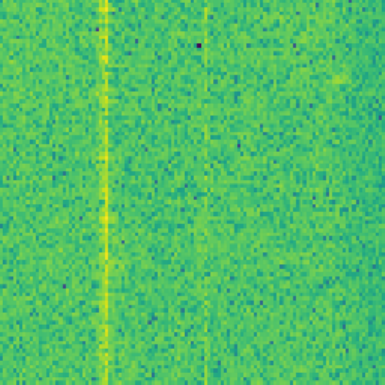}
    \caption{Visibility data}
    \label{fig:11}
  \end{subfigure}\hspace{1em}
   \begin{subfigure}[t]{0.18\textwidth}
    \includegraphics[width=\textwidth]{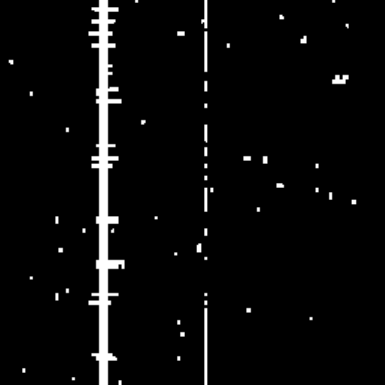}
    \caption{Ours mask}
    \label{fig:12}
  \end{subfigure}\hspace{1em}
   \begin{subfigure}[t]{0.18\textwidth}
    \includegraphics[width=\textwidth]{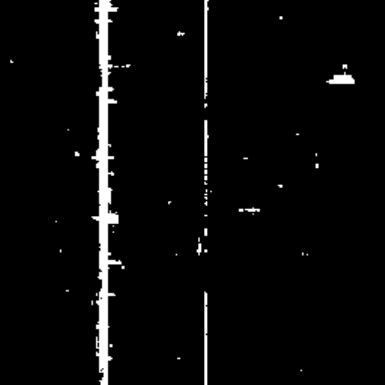}
    \caption{R-Net mask}
    \label{fig:13}
  \end{subfigure}\hspace{1em}
   \begin{subfigure}[t]{0.18\textwidth}
    \includegraphics[width=\textwidth]{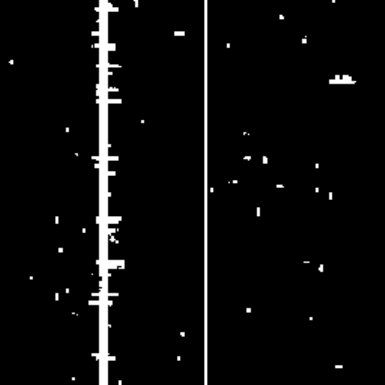}
    \caption{RFI-Net mask}
    \label{fig:14}
  \end{subfigure}\hspace{1em}
   \begin{subfigure}[t]{0.18\textwidth}
    \includegraphics[width=\textwidth]{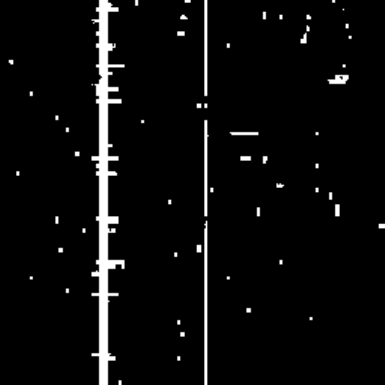}
    \caption{U-Net mask}
    \label{fig:15}
  \end{subfigure}\hspace{1em}
   \begin{subfigure}[t]{0.18\textwidth}
    \includegraphics[width=\textwidth]{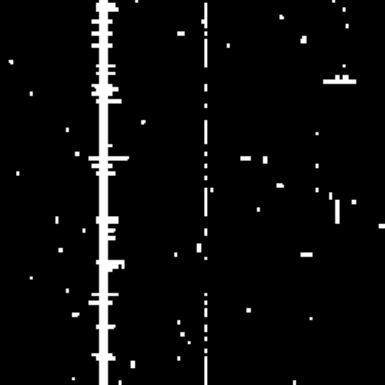}
    \caption{Ground truth}
    \label{fig:16}
  \end{subfigure}\hspace{1em}
   \begin{subfigure}[t]{0.18\textwidth}
    \includegraphics[width=\textwidth]{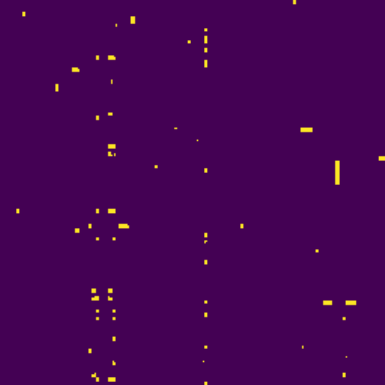}
    \caption{Ours difference map}
    \label{fig:17}
  \end{subfigure}\hspace{1em}
   \begin{subfigure}[t]{0.18\textwidth}
    \includegraphics[width=\textwidth]{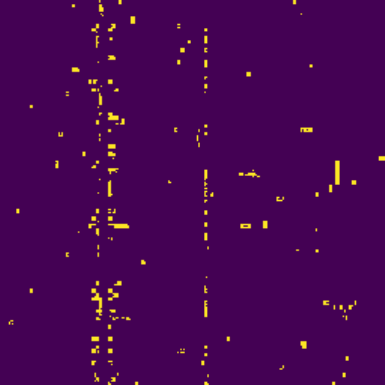}
    \caption{R-Net difference map}
    \label{fig:18}
  \end{subfigure}\hspace{1em}
   \begin{subfigure}[t]{0.18\textwidth}
    \includegraphics[width=\textwidth]{img/result/r_net_difference_map_67.pdf}
    \caption{RFI-Net difference map}
    \label{fig:19}
  \end{subfigure}\hspace{1em}
   \begin{subfigure}[t]{0.18\textwidth}
    \includegraphics[width=\textwidth]{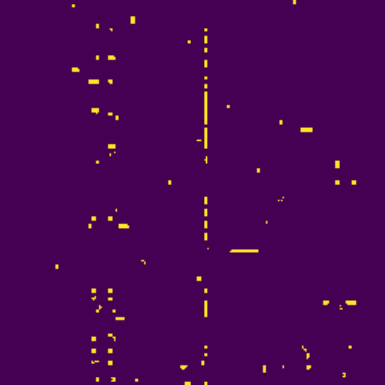}
    \caption{U-Net difference map}
    \label{fig:20}
  \end{subfigure}\hspace{1em}
  \caption{Visualization of experimental results. In each image, the horizontal axis is the number of frequency channels, representing the frequency axis, and the vertical axis is the number of subintegrations, representing the time axis. The green background in the first column on the left represents the visibility data, the black background represents the ground truth corresponding to the visibility data, and the white area represents the presence of RFI. The second to last columns of the first and third rows show the segmentation masks output by different models. Likewise, white areas represent model predictions of the presence of RFI. The second to last columns of the second and fourth rows present the difference maps between the segmentation masks output by different models and the ground truth. This figure is obtained by subtracting the absolute value of the segmentation mask output from the model and the ground truth. The yellow area indicates a difference.}

  \label{fig:vision result}
\end{figure*}

\subsection{Performance evaluation}
\label{sec:Performance evaluation}
\begin{table}
	\centering
	\caption{Detailed performance metric scores of the different models on observational data from the Yunnan Observatory's 40-meter radio telescope, using data augmentation. The highest score on each metric is shown in bold.}
	\label{tab:using augmentation}
	\begin{tabular}{lcccr} 
		\hline
		Model & Precision & Recall & F1 score & Iou\\
		\hline
		EMSCA-UNet(ours) & \textbf{88.08} & \textbf{83.62} & \textbf{85.80} & \textbf{75.54}\\
		U-Net& 83.40 & 78.25 & 80.74& 69.28\\
		RFI-Net& 83.72 & 78.90 & 81.24& 69.41\\
            R-Net& 78.12 & 63.57 & 70.01& 55.54\\
		\hline
	\end{tabular}
\end{table}

\begin{table}
	\centering
	\caption{Detailed performance metric scores of the different models on observational data from the Yunnan Observatory's 40-meter radio telescope, without using data augmentation. The highest score on each metric is shown in bold.}
	\label{tab:without augmentation}
	\begin{tabular}{lcccr} 
		\hline
		Model & Precision & Recall & F1 score & Iou\\
		\hline
		EMSCA-UNet(ours) & \textbf{87.73} & 79.78 & \textbf{83.57} & \textbf{72.60}\\
		U-Net& 81.62 & 74.45 & 77.87& 65.92\\
		RFI-Net& 79.47 & \textbf{80.39} &79.93& 67.77\\
            R-Net& 75.67 & 66.74 & 70.93& 56.47\\
		\hline
	\end{tabular}
\end{table}

\begin{table}
	\centering
	\caption{The first two rows of the table present the results of ablation studies carried out on the MSCA and ECA modules."Only MSCA Block"signifies that the ECA Block was removed from our model architecture, rendering the two convolutional layers of U-Net functionally equivalent to the MSCA Block. Similarly, the term "Only ECA Block" denotes that the first convolutional layer of U-Net was replaced with the ECA Block, leaving the second convolutional layer unchanged. The third row provides detailed scores obtained by our model utilizing the Adam optimizer, serving to compare the differences between Adam and AdamW and to substantiate the rationality of employing AdamW.}
	\label{tab:ablation}
	\begin{tabular}{lcccr} 
		\hline
		Model & Precision & Recall & F1 score & Iou\\
		\hline
		Only MSCA Block & 87.92 & 81.93 & 84.82 & 74.51\\
		Only ECA Block& 85.92 & 80.63 & 83.19& 71.62\\
		EMSCA(ours) with Adam & 87.39 & 83.94 &85.63& 75.35\\
		\hline
	\end{tabular}
\end{table}

The detection results of different models on two different datasets are shown in Fig.~\ref{fig:vision result}. The left column of the figure displays the Visibility data and their corresponding Ground truth. The second to last columns depict the flagging results of different models, as well as the difference maps between the model outputs and the Ground truth. From the figure, it can be observed that our model achieves the best flagging performance, with the output masks being closest to the Ground truth. This indicates finer edge detection, lower false detection rate for RFI, and fewer false positives. The difference maps are intended to provide a more intuitive visualization of the discrepancies between the output masks of different models and the Ground truth. By calculating the difference between the Ground truth and the output masks of different models and taking the absolute value, the yellow areas in the maps represent the differences between the predicted masks and the Ground truth. It can be clearly observed from these difference maps that our model has the fewest yellow error regions and the smallest differences with the Ground truth. Table~\ref{tab:using augmentation} presents the scores of four evaluation metrics (precision, recall, F1 score, and IoU) achieved by our EMSCA-UNet model, U-Net model, RFI-Net model, and R-Net model on the test set. To facilitate understanding, the highest score for each metric is highlighted in bold. Specifically, our EMSCA-UNet model achieves scores of 88.08, 83.62, 85.80, and 75.54 for precision, recall, F1 score, and IoU, respectively. The U-Net model and the RFI-Net model obtain similar scores across these four metrics, but the RFI-Net model slightly outperforms the U-Net model in each metric. However, the R-Net model performs poorly, with scores lower than other models in each metric. Overall, our model achieved the highest scores on all four evaluation metrics, averaging about a 5\% improvement over the U-Net model. The performance of RFI-Net is comparable to U-Net, with RFI-Net slightly outperforming U-Net, while the performance of R-Net is the worst.
\subsection{Comparative analysis and discussion}
\label{sec:Comparative analysis and discussion}
Table~\ref{tab:without augmentation} displays the specific scores of each model without the use of data augmentation. Upon comparing Tables~\ref{tab:using augmentation} and~\ref{tab:without augmentation}, a slight decline in most of the model's performance metrics can be observed when data augmentation is not applied. This observation substantiates the effectiveness of our implemented data augmentation method in enhancing the model's generalization capabilities.
By comparing the structures of R-Net with the other three models, it can be observed that our EMSCA-UNet and RFI-Net models adopt the U-net model's U-shaped encoder-decoder architecture and employ multiple skip connections. In contrast, R-Net utilizes the classic residual network structure (ResNet) but only employs a single skip connection. Aside from R-Net, the three other models have different numbers of RFI feature maps extracted by the convolutional layers throughout the entire input-to-output process. In the encoder phase, the number of feature maps gradually increases, reaching up to a maximum of 1024. In the decoder phase, the number of feature maps gradually decreases. Different from the other three models, each layer of R-Net only has a fixed number of 12 feature maps. Consequently, through experimental results, it can be concluded that skip connections and a greater number of feature maps are crucial for our RFI detection task. This further corroborates our previous statement in Section ~\ref{sec:Architecture of EMSCA-UNet}, emphasizing the necessity of integrating skip connections to mitigate network degradation and information loss caused by downsampling, while more feature maps enable our model to learn a wider range of features. By comparing our EMSCA-Unet and RFI-Net models to the U-Net model, it can be observed that RFI-Net augments U-Net by adding an additional 3×3 convolutional layer and introducing Shortcut Connections at each stage of the encoder and decoder. Hence, RFI-Net and U-Net share a striking structural resemblance, explaining the close experimental results between the two. Our model effectively replaces the first of the two 3×3 convolutional layers in U-Net with an ECA module, and the second with an MSCA module, incorporating Shortcut Connections in the MSCA module. Therefore, in a sense, our comparative study between EMSCA-UNet and U-Net can be considered an ablation study. Comparing the first and second rows of Table~\ref{tab:ablation} with the first row of Table~\ref{tab:using augmentation}, it is evident that the performance of individual MSCA blocks or ECA blocks is inferior to when both are used together. Furthermore, from the first and second rows of Table~\ref{tab:ablation} and the second row of Table~\ref{tab:using augmentation}, it can be observed that replacing both convolutional layers of U-Net with MSCA block or substituting the first convolutional layer with ECA block leads to improved performance compared to the original U-Net. These findings indicate the effectiveness of both modules added, and their combined usage yields better results than individual usage. The use of the Efficient Channel Attention (ECA) module allows our model to focus more on feature maps that are pertinent to RFI detection during the training process. Moreover, by utilizing the Multi-Scale Convolution Attention (MSCA) module, which extracts multi-scale features, the model effectively enhances RFI detection compared to solely adopting single-scale 3×3 convolutions. The subpar performance of R-Net may be attributed to the need for transfer learning, as outlined in their original paper where a significant performance improvement was achieved by training on simulated data and fine-tuning on real data. In addition the third row of Table~\ref{tab:ablation} shows the experimental results obtained for our model using the adam optimizer, while the experimental results in Table~\ref{tab:using augmentation} are all obtained using the adamW optimizer. By comparing the results with the first row of Table~\ref{tab:using augmentation} it can be seen that the experimental results obtained using adamW and adam are very close, but using adamW leads to a little very small improvement in the model performance. Although this improvement is very small it shows that it is reasonable to use adamW as our optimizer. Despite our model's superiority in detection results and various metrics, it does not exhibit an advantage in terms of speed, which is an aspect that can be further optimized in our future work.

\section{CONCLUSION}
\label{sec:conclusion}

The presence of RFI significantly affects the quality of radio astronomical observation data. Consequently, accurately detecting RFI from radio observation data is of paramount importance. Currently, the application of several state-of-the-art supervised deep learning techniques in detecting RFI in the observational data of the Yunnan Observatory's 40-meter radio telescope is not yet effective enough, and there are often problems of misdetection and omission. 

In this study, we propose a novel EMSCA-UNet model that combines the advantages of convolutional operations and attention mechanisms. We conduct experiments using the observation data from the 40-meter radio telescope at Yunnan Observatory and demonstrate that, compared to several state-of-the-art supervised RFI detection methods, EMSCA-UNet exhibits superior performance in RFI identification. It achieves more comprehensive RFI detection, higher precision, lower error rates, and finer marginal detail. However, we acknowledge limitations in our approach. Firstly, it involves the labeling issue faced by almost all supervised deep learning methods, as the quality of labels largely influences the performance of the final model. Currently, we address this issue by manually verifying the correctness of labels. Secondly, the efficiency of our method in terms of computational speed is not yet optimal. In future work, we will continue to optimize our model and explore techniques such as pruning or parallel computing to improve computational speed. Furthermore, in this paper, we explore the application of attention mechanisms in the detection of RFI and find that attention mechanisms still have untapped potential and wide application space in RFI detection. This prompts us to further explore and develop more possibilities in the identification and handling of RFI.

\section*{Acknowledgements}

We are very grateful to the reviewers for their valuable comments. This work was supported by the National Key Research and Development Program of China (2020SKA0110300, 2020SKA0120100), and National Natural Science Foundation of China (12063003, 12073076). The authors acknowledge financial support from the Yunnan Ten Thousand Talents Plan Young \& Elite Talents Project.The author is very grateful to Chris Finaly for his insightful discussions and comments.

\section*{Data Availability}
The data will be made available on reasonable request from the authors.



\bibliographystyle{mnras}
\bibliography{example} 




\appendix

\section{METRICS}
\label{sec:appendix}

\begin{equation}
Precision=\frac{True \ Positive}{True \ Positive+False \ Positive} 
	\label{eq:precision}
\end{equation}
Precision measures the ability of our model to correctly identify RFI within the instances that it has flagged as such. A higher precision means that the model has fewer false positives (mistakenly identified as RFI) when detecting RFI.

\begin{equation}
Recall=\frac{True \ Positive}{True \ Positive+False \ Negative} 
	\label{eq:recall}
\end{equation}
Recall (Recall) reflects the proportion of actual RFI that the model can detect for all RFI. A high recall rate means that the model has fewer false negatives (missed RFI) when detecting.

\begin{equation}
F1=\frac{2\times Precision\times Recall}{Precision+  Recall} 
	\label{eq:f1}
\end{equation}
The F1 Score is the harmonic mean of precision and recall, used to balance these two measures. When either precision or recall is low, the F1 Score will also decrease, therefore reflecting that the model needs to perform well in both aspects.

\begin{equation}
Iou=\frac{True \ Positive}{True \ Positive + False \ Positive + False \ Negative} 
	\label{eq:iou}
\end{equation}
The Intersection over Union (IoU) represents the degree of overlap between the model's predicted segmentation area and the actual ground truth segmentation area. It indicates that the closer the model's predicted segmentation is to the actual segmentation, the better the segmentation effect. Therefore, using IoU as our evaluation metric in our RFI detection task can provide a more intuitive understanding of the quality of the detection results.


\bsp	
\label{lastpage}
\end{document}